\documentclass[runningheads]{llncs}
\usepackage[T1]{fontenc}
\usepackage[misc]{ifsym}
\usepackage{graphicx}
\usepackage{color}
\usepackage{colortbl}
\RequirePackage{cite}
\usepackage{amssymb}
\usepackage{graphicx}
\usepackage{caption}
\usepackage{tcolorbox}
\usepackage{adjustbox}
\usepackage{graphicx}
\usepackage{algorithm}
\usepackage{algpseudocode}
\RequirePackage{algorithm}

\usepackage{multirow}
\usepackage[utf8]{inputenc}
\usepackage{xcolor}
\usepackage[mathscr]{euscript}
\RequirePackage{amsmath,amssymb,bm}
\usepackage{paralist}
\usepackage{nameref}
\usepackage[textsize=scriptsize,color=yellow!25]{todonotes}
\newcolumntype{P}[1]{>{\centering\arraybackslash}p{#1}}

\newcommand{\mlp}{\textsc{Mlp}}
\newcommand{\sbert}{\textsc{Sbert}}
\newcommand{\gat}{\textsc{Gat}}
\newcommand{\gcn}{\textsc{Gcn}}
\newcommand{\sage}{\textsc{GraphSage}}
\begin{document}
\title{Is this bug severe? A text-cum-graph\\ based model for bug severity prediction}
%
\author{Rima Hazra\inst{1, 2} \Letter\and
Arpit Dwivedi\inst{1, 2} \and
Animesh Mukherjee\inst{1, 3}}
\authorrunning{Hazra et al.}
%
\institute{Indian Institute of Technology Kharagpur, India\\ \and
\email{\{to\_rima, arpitdwivedi\}@iitkgp.ac.in} \and
\email{\{animeshm\}@cse.iitkgp.ac.in}
}
\toctitle{Is this bug severe? A text-cum-graph\\ based model for bug severity prediction}
\tocauthor{Rima~Hazra}
\maketitle              

\begin{abstract}
Repositories of large software systems have become commonplace. This massive expansion has resulted in the emergence of various problems in these software platforms including identification of (i) bug-prone packages, (ii) critical bugs, and (iii) severity of bugs. One of the important goals would be to mine these bugs and recommend them to the developers to resolve them. The first step to this is that one has to accurately detect the extent of severity of the bugs. In this paper, we take up this task of predicting the severity of bugs in the near future. Contextualized neural models built on the text description of a bug and the user comments about the bug help to achieve reasonably good performance. Further information on how the bugs are related to each other in terms of the ways they affect packages can be summarised in the form of a graph and used along with the text to get additional benefits.
\end{abstract}
\section{Introduction}
\label{sec:intro}
Large software systems have become increasingly commonplace. As these repositories grow, they become more and more complex. Malfunctions (aka bugs) in such systems need to be tackled in a timely fashion. Such bugs can be reported by the end-users (who are using the service), the developers or the testers. Since these bugs need to be attended in a pipelined fashion, an important step is to understand and prioritize the bug reports as per their severity. For instance, the security related bug reports should possibly get more priority than any other types of reports. Here, the term ``severity" corresponds to the important bugs. The importance can be of security, privacy, affecting users etc. Though the actual definition of "severity" and "priority" from a business perspective is different, the term "severity" has been used here to represent the critical bugs. Since the number of bugs in a large software system could be in millions, it is difficult for the developers to manually go through the list of bug reports and identify the most important bug from that list. Not only is this a very tedious task but also is prone to mistakes. Thus automatic methods to predict the severity of a bug is very crucial. Such models can predict the bugs that are going to soon become severe based on certain early indicators like the description and the comments about the bug plus the number of packages that it already affects. 

\noindent{\bf Bug severity prediction} is a task of predicting the severity/impact of each bug from a huge list of bugs. Bug severity prediction task has been performed on many software platforms such as mozilla\footnote{\url{https://bugzilla.mozilla.org/home}}, eclipse\footnote{ \url{https://bugs.eclipse.org/bugs/}}, GCC\footnote{\url{https://gcc.gnu.org/bugzilla/}} etc. by past researchers~\cite{Umer:2018,Tan:2020,Arokiam:2020,Wu:2021}. Most earlier models for bug severity detection assumes it to be a classification problem whereby the task is to predict one of the severity states -- critical, major, normal, minor, trivial, enhancement. None of the datasets contains actual parameter based scores of the bugs. The other issue is that none of these models leverage the usefulness of both text and graph information jointly.
Even while the state of a bug is predicted by earlier models, there might still be many bugs that need to be manually sorted within each class. Hence in this paper, we present a regression model that generates the rank of each bug in the list, allowing the developers to set their priorities accurately. To this purpose, we curate a new dataset consisting of bugs mapped to the Ubuntu packages that affect those packages over time. This new dataset allows us to perform the experiments in the regression setup. Further, unlike earlier models, we use sophisticated neural architectures that suitably blend text and graph information to harvest the benefits from both of these information sources. \\
\noindent\textbf{Our contributions and results ---} \\
\noindent\textbf{{\em A new dataset}}: We curate a new dataset~\footnote{\url{https://doi.org/10.5281/zenodo.5554974}} consisting of $\sim$280K bugs along with their meta data (e.g., the textual description of the bug, etc.). Further the ground-truth severity scores of these bugs have been collected in two different time points to facilitate the prediction experiments. In addition, we have also curated the list of packages affected by these bugs over time.\\
\noindent\textbf{{\em Bug severity prediction}}: As noted earlier we perform the experiments in the regression setup. In particular, we develop different neural models based on the text obtained from the bug description and user/developer/maintainer comments. We also use graph information in the form of how the bugs are related to each other in terms of the packages they co-affect. This is one of the most unique points of our approach. Of particular interest is the observation that the use of the text-cum-graph based models outperform the text based models in low training data settings. A summary of the key results are as follows -- with 70\% training data the most competing text based method \sbert~outperforms the text-cum-graph based methods like \gat~and \gcn. However, in this setting \sage~performs the best. For the low data setting ($\le 25\%$ training data), the text-cum-graph based models largely outperform the purely text based models. Notably, for 5\% training data, while \sbert~achieves an MAE, MSE and MAPE of 1.178, 2.268 and 0.1778 respectively while \gat ~reports an MAE, MSE and MAPE of 0.849, 1.415 and 0.1101 respectively.
\section{Related Work}
\label{sec:related_work}
Due to the large-scale of software project, many significant problems require automated systems. Some of the major problems of large-scale software systems include identifying high impact bugs (security bug reports), recommending appropriate developers to packages (/modules), identifying bug-prone packages, retrieving duplicate bugs, predicting high priority bugs, predicting severe bugs. Most of these problems were/are handled manually by developers/maintainers in the past, but as the data increases, the automated system is the need of the hour. Some of these problems have been studied for a long time. Researchers have used various metadata, textual information, underlying graph structures to build systems that can solve certain problems. There are very few datasets available for each of the problems. In this section, first, we will discuss the datasets available, then we briefly discuss the various earlier proposed methods into two parts -- text based methods and graph based methods. 
For high impact bug reports (security bug prediction), there are four datasets -- Ambari, Camel, Derby, and Wicket had been first manually labelled by ~\cite{Ohira:2015} and then further relabelled (only mislabelled data) by ~\cite{Wu:2020,Peters:2019}. Two large scale projects -- Chromium, and OpenStack datasets\footnote{\url{https://github.com/wuxiaoxue/cve-assisted}} were constructed by Wu {\em et al.}~\cite{Wu:2020}.
In~\cite{Tan:2020}, the authors collected bug reports and their severity levels from GCC, OpenOffice, Eclipse, NetBeans, and Mozilla for bug severity prediction. For predicting the severity of bug reports, Ramay {\em et al.}~\cite{Ramay:2019} used the bug reports of seven open-source products -- Platform, CDT, JDT, Core, Firefox, Thunderbird, and Bugzilla from the repository created by ~\cite{Lamkanfi:2013}. So, the datasets available for severity prediction are mostly around these open-source projects, and the severity has only the label but not the scores.
For most of the problems~\cite{Arokiam:2020,Wu:2021,Peters:2019}, researchers have used textual information such as title, descriptions to solve problems like bug severity prediction. In different papers~\cite{Peters:2019, Ramay:2019, Arokiam:2020}, authors have used TF-IDF, word2vec~\cite{mikolov:2013}, glove embeddings, doc2vec, BM25 to represent the text. 
The bug severity prediction is tackled as a classification problem in earlier works~\cite{Goseva-Popstojanova:2018, Wu:2021, Arokiam:2020}. Authors have used various classification algorithms like SVM, logistic regression, random forest and XGboost to solve the problem. These studies have been performed on different platforms like Mozilla, eclipse, GCC etc. In~\cite{Tan:2020}, the authors proposed a method based on logistic regression to predict the severity of bugs. They used data from all the above three platforms for their experiments. Ramay {\em et al.}~\cite{Ramay:2019} proposed a deep learning approach to predict the severity of software bugs based on the text present in the bug report. 
Umer {\em et al}~\cite{Umer:2018} proposed an emotion-based approach to predict the priority of bug reports.
There is very little research on the utility of graphs in the software domain to solve major software system problems. In~\cite{Hazra:2021}, the authors studied in detail the problem of bug urgency ranking and developer recommendation with the help of underlying graph structures. The authors curated a dataset from the Ubuntu platform for their experiments and developed various machine learning models for the ranking and recommendation problems. 
Our work is unique in two different ways. The first and possibly simple point of difference is that we formulate the problem in a regression setup and curate a new dataset for this purpose. The second and the most unique point is that we propose a novel graph formulation of the problem that allows us to do reasonably good predictions even when very low training data points are available.
\section{Dataset}
\label{sec:datasets}
\begin{figure*}
\begin{minipage}{.33\textwidth}
\centering
\includegraphics[width=\textwidth]{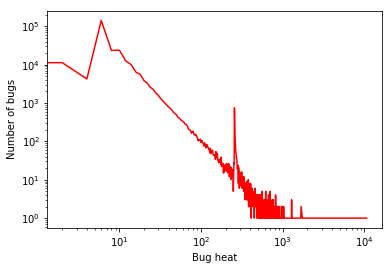}
\caption{\label{fig:bug_distribution}Distribution of bugs having a specific bug heat.}  
\end{minipage}
\begin{minipage}{.33\textwidth}
\includegraphics[width=\textwidth]{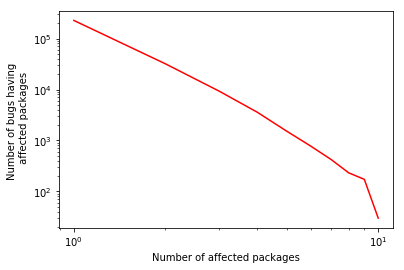} \\
\caption{Distribution of bugs affecting a given number of packages.}  
\label{fig:bug_affects}
\end{minipage}
\begin{minipage}{.32\textwidth}
\includegraphics[width=\textwidth]{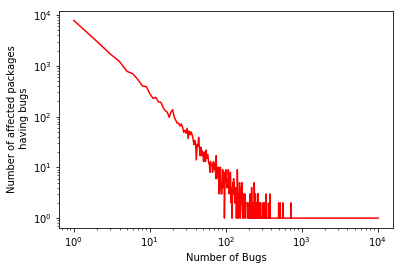} \\
\caption{Distribution of packages affected by a given number of bugs.}  
\label{fig:package_affected}
\end{minipage}
\end{figure*}

\begin{figure*}
\begin{minipage}{1\textwidth}
\includegraphics[width=1\textwidth]{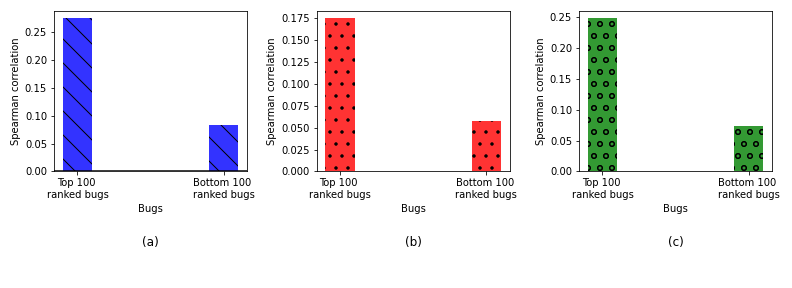}
\vspace{-1.4 cm}
\caption{(a) Spearman's rank correlation between the (a) degree centrality, (b) clustering coefficient and (c) PageRank and bug heat ranks for top 100 ranked bugs and bottom 100 ranked bugs based on bug heat.} 
\label{fig:networkF}
\end{minipage}
\end{figure*}

We collect $\sim$280K bug reports related to Ubuntu repositories reported within the time span of 2004 to 2019. These bug reports are collected from the launchpad\footnote{https://launchpad.net/} bug tracking system. Each bug page consists of various meta information such as the title, the description, the name of the bug reporter, reporting timestamp, comments with timestamp, activity log\footnote{https://bugs.launchpad.net/ubuntu/+source/linux/+bug/1945590/+activity}, packages affected by the bug with timestamp and the bug heat~\footnote{\url{https://bugs.launchpad.net/+help-bugs/bug-heat.html}} (or severity). Bug description consists of textual information written by the bug reporter. Given a bug, activity log keeps track of each and every activity that are made on the bug. `Affected' packages are the packages that were affected by the given bug at a given time point. The bug heat is the accumulated score based on factors like privacy issue, security issue, duplicate nature, affected users and subscribers. This bug heat score is a representative of the severity/urgency of a bug. The bug heat calculation score is given in Table~\ref{tab:bugheat}. An example of a bug entry is given in Table~\ref{tab:BugDetails}. 
\begin{table*}[!ht]
\tiny
\scalebox{0.9}{
\begin{tabular}{|p{2.5cm}|p{10.5cm}|} \hline 
{\bf Field} & {\bf Information} \\ 
  \hline \hline
Bug Id & 1663552\\ \hline
Reported On & mysql-5.7 \\ \hline
Description & Hi, on one of our servers we noticed that under certain conditions mysql-server can be caused to go berserk, i.e. run with 400\% CPU load, spit out extrem tons of log messages and denial it's work completely when contacted by a client, that is not (!) authorized to connect... \\ \hline
Comments & {\bf [2017-02-10 20:37:45 UTC]} Thanks for the bug user; I'm marking this public so that administrators can more quickly learn that using tcpwrappers for access control has the potential ... \\
& {\bf[2017-03-06 11:19:44 UTC]} user, could you report the package version number of mysql-5.7 in which you are seeing this please? ...\\
& {\bf[2017-03-06 23:22:02 UTC]} user, I did not keep the virtual machine. On a host where the problem occured first we have ...\\ \hline
Affected packages & {\bf [10-02-2017]}mysql-5.7 \\
&{\bf [30-03-2017]} mysql server \\ \hline
Bug heat score & 16\\ \hline
\end{tabular}
}
\caption{\label{tab:BugDetails} Released metadata of a bug.}
\end{table*}
High bug heat represents a more severe bug. In our experiment, we have considered only those bugs which have at least one comment. We have collected the bug heat of the bugs at two different time points -- in November 2019 and again in November 2020. Table~\ref{tab:datastat} notes the basic statistics of the data collected. In Figure~\ref{fig:bug_distribution}, we show the distribution of bugs having a particular bug heat value. The distribution is highly skewed with most bugs having low severity and a few bugs with very high severity.\\
\noindent \textbf{Bug-package network}: From our dataset, we can conceive a bipartite network where one set contains the list of bugs ($B$) and the other set contains the list of packages ($P$). A bug $b \in B$ can have a directed edge to a package $p \in P$ if the bug $b$ affects the package $p$. An example of such a bipartite network shown in Figure~\ref{fig:example_graph}. Given a bug, affected packages are added typically along with a timestamp. However, in a few cases, this timestamp is not available. In such cases we replace the unavailable timestamp with the time of the bug creation assuming that the package is being affected since the creation of the bug report.
\begin{table*}[!htb]
\centering
\tiny
\begin{minipage}{0.4\textwidth}
\begin{tabular}{|p{2cm}|p{2.5cm}|} \hline
{\bf Attribute} & {\bf Calculation}  \\ \hline \hline
Private & adds 150 points\\ \hline
Security issue & adds 250 points\\ \hline
Duplicates & 6 points per duplicate bug \\ \hline
Affected users & 4 points per affected user\\ \hline
Subscribers~\footnote{(incl. subscribers to duplicates)}  & 2 points per subscriber \\ \hline
\end{tabular}
\caption{Bug heat score calculation strategy}\label{tab:bugheat}
\end{minipage}
\begin{minipage}{.45\textwidth}
\begin{tabular}{|p{4.9cm}|p{1.2cm}|} \hline
{\bf Basic information} & {\bf Count}  \\ \hline \hline
 Total number of bugs & 273,544\\ \hline
Average number of comments & 5.241\\ \hline
Average number of words in description & 80.93 \\ \hline
Maximum number of words in description & 4967\\ \hline
Average number of words in comments & 10.959 \\ \hline
Maximum number of words in comments & 436\\ \hline
Average number of affected packages & 1.28\\ \hline
\end{tabular}
\caption{Dataset statistics.}\label{tab:datastat}
\end{minipage}
\end{table*}
\noindent\textit{Degree distribution}: In Figure~\ref{fig:bug_affects}, we illustrate the distribution of bugs affecting the different number of packages while in Figure~\ref{fig:package_affected} we show the distribution of packages affected by the different number of bugs. Both these distributions exhibit a scale-free behaviour.\\
\begin{figure*}
\centering
\begin{minipage}{.6\textwidth}
\resizebox{1\textwidth}{!}{
\begin{tikzpicture}
  \node (s) [draw=red, fill=gray!30] {45702};
  \node (s1) [right=1cm of s, draw=red, fill=gray!30] {64371};
  \node (s2) [right=1cm of s1,  draw=red, fill=gray!30] {1022921};
  \node (s3) [right=1cm of s2,  draw=red, fill=gray!30] {1566870};
  \node (s4) [right= 1cm of s3,  draw=red, fill=gray!30] {1298939};
  \node (d1) [above = 15mm of s1, draw=red, fill=red!10] {mono};
  \node (d2) [right=1cm of d1,draw=red, fill=red!10] {banshee};
  \node (d3) [right = 1cm of d2,draw=red, fill=red!10] {rhythmbox};
  
  \draw[->] (s) edge [draw, dashed, thick] (d1);
  \draw[->] (s1) edge [draw, dashed, thick] (d2);
  \draw[->] (s1) edge [draw, dashed, thick] (d1);
  \draw[->] (s2) edge [draw, dashed, thick] (d2);
  \draw[->] (s3) edge [draw, dashed, thick] (d2);
  \draw[->] (s3) edge [draw, dashed, thick] (d3);
  \draw[->] (s4) edge [draw, dashed, thick] (d3);
\end{tikzpicture}
}
\caption{Bipartite network between set of bugs and set of packages. $B=$\{45702, 64371, 1022921, 1566870 and 1298939\}. $P=$\{`mono', `banshee', `rhythmbox'\} are the affected packages.}
\label{fig:example_graph}
\end{minipage}
\begin{minipage}{.3\textwidth}
\centering\scriptsize
\resizebox{\textwidth}{!}{
\begin{tikzpicture}
  \node (s) [draw=red, fill=gray!30] {45702};
  \node (s1) [above = 7 mm of s, draw=red, fill=gray!30] {64371};
  \node (s2) [above left=1cm of s,  draw=red, fill=gray!30] {1022921};
  \node (s3) [left=1cm of s,  draw=red, fill=gray!30] {1566870};
  \node (s4) [below right = 7 mm of s3,  draw=red, fill=gray!30] {1298939};
  \draw[-] (s) edge [draw, thick] (s1);
  \draw[-] (s1) edge [draw, thick] (s2);
  \draw[-] (s1) edge [draw, thick] (s3);
  \draw[-] (s2) edge [draw, thick] (s3);
  \draw[-] (s3) edge [draw, thick] (s4);
\end{tikzpicture}
}
\caption{Bug-bug network.}
\label{fig:projected_graph}
\end{minipage}
\end{figure*}
\noindent \textbf{Bug-bug network}: From the bug-package bipartite network we construct an one-mode projection, i.e., a bug-bug network where $B$ is the set of nodes in this network and two nodes $b_i$ and $b_j$ are connected if they co-affect a package. Figure~\ref{fig:projected_graph} shows the corresponding of bug-bug network constructed from Figure~\ref{fig:example_graph}. For instance, in this figure bug 45702 and 64371 are connected because both of them affect the same package `mono'. We shall use the information from this network for all our learning algorithms in the subsequent sections.\\
To understand how the bug-bug network structure is correlated with the bug heat ranks, we analyze a few known network properties of 100 top and bottom-ranked bugs based on bug heat rank. First, we compute three node centric measures: degree centrality, clustering coefficient, and PageRank. We have chosen these measures because they will give us an idea of the neighbourhood quality (i.e., how dense it is and how many neighbours a node has). Given a node centrality measure, we rerank the bugs based on these measures and compute the Spearman's rank correlation with the actual 100 top-ranked and bottom-ranked bugs based on bug heat. In Figure~\ref{fig:networkF}, we plot the correlation values for the top 100 and bottom 100 bugs for all three measures. In all cases we observe that the top ranked bugs are more strongly correlated with the network properties. This gives us the first indication that it is important to leverage the network structure in order to efficiently perform bug severity prediction.

\section{Bug severity prediction}
Given a set of bugs ($B$), our objective is to predict their ranks based on the bug heat score. In particular, as is usual in a regression setup we are interested to predict the $\log(rank(bug\textrm{-}heat))$. We do not directly predict the bug heat since the distribution is very skewed (see Figure~\ref{fig:bug_distribution}).
To begin with every bug is encoded as a combination of the text description and the user comments. Often, the description of the bug contains additional information like stack traces, code fragments email ids and urls. We remove these pieces of information using some simple heuristics.\\ 
\noindent \textbf{Text based approach}: In case of text based approach, we primarily use the textual description and the user comments for a bug for the purpose of prediction. \\
\noindent \textbf{Doc2Vec}: Doc2Vec~\cite{quoc:2014} algorithm is used to generate representation for a document. It follows the same architecture as word2vec~\cite{mikolov:2013} along with a new vector called paragraph id. This paragraph id is used to represent each document uniquely. While training, along with the word vectors, a document vector is also trained and at the end it generates the representation of documents. We train the doc2vec~\cite{quoc:2014} model on our corpus. We construct 100 dimension embeddings for the descriptions and the comments separately. Finally, a bug is a concatenation of these two vectors. This is passed through an \mlp~to obtain the regression scores. The \mlp~has one hidden layer followed by the output layer and the activation function is \textsc{RelU}.\\
\noindent \textbf{SBERT}: SBERT~\cite{reimers:2019} has the BERT~\cite{Devlin:2019} like architecture which can capture better semantics in the sentence. It is the fine tuned BERT sentence embedding model which can correctly capture the the semantic textual similarity (STS) between a pair of sentences. We use the pretrained sentence BERT~\cite{reimers:2019} model (\sbert) to generate embeddings. For each bug, we pass the preprocessed description text and the comment text through the model and obtain separate embeddings for each of these. We then concatenate these two embeddings and pass it through the same \mlp~model discussed earlier to obtain the regression scores.\\
\noindent \textbf{Graph based approaches}:\\
\noindent \textbf{GAT}: Graph attention network (GAT)~\cite{velickovic:2018} algorithm can again be used for node classification where the input features are linearly transformed to some output features which is further followed by a self-attention layer on the nodes. This self-attention mechanism captures the importance of one node on another. For our purpose once again the bug-bug network is used as the input graph and each node is initially encoded as a concatenated vector of the \sbert~representations of the bug description and the comments. Finally, the regression scores are once again obtained using a linear layer and a \textsc{RelU} activation. Here again the model is only shown the labels of the training nodes.\\
\noindent \textbf{GCN}: Graph convoluation network (\gcn)~\cite{kipf:2017} is a transductive method for node classification. For our purpose we have used the bug-bug network as the input graph. To begin with each node is featurised as a concatenation of the \sbert~representations of the description and the comments. Finally, as earlier, to obtain the regression scores, we pass the output features of the GCN layer to a linear layer with \textsc{RelU} activation. Like before, we show the model only the labels of the training nodes. \\
\noindent \textbf{GRAPHSAGE}: \sage~\cite{hamilton:2017} is a graph based algorithm for classifying nodes. In this method, neighbourhood aggregation has been done using uniformly sampled neighbours and aggregating their features. In our case we use the bug-bug network and feed it to \sage. The initial representation of every bug (read node) is a concatenated vector of the \sbert~embeddings of the description and the comments. For the purpose of training we provide the label of the nodes (i.e., their $\log(rank(bug\textrm{-}heat)$) that are part of the training set only (we shall discuss more about this in the following section on experimental setup). We obtain the final predictions using a linear layer with \textsc{RelU} activation.
\section{Experiments and results}
\label{sec:experimental_setup}
\noindent \textbf{Experimental setup}\\
\noindent \textit{Training}: In order to train the models, we consider all the bugs that have been posted in between January 2017 -- June 2017. We have considered only the bugs which has at least one comment. This results in a total of 5835 bugs. Given a bug in this time period, all comments posted about the bug in between January 2017 -- June 2018 are considered to compute the comment based embedding. This ensures that each bug has at least one year span of comments for the computation of the features. The ground truth bug heat scores that we use have been crawled in November 2019. Out of the 5835 bugs obtained, we use different training set sizes for our experiments ranging from 70\% to 5\%. The rest of the data is used for validation in each case. For each of the training and the validation sets the bugs selected are re-ranked within that set based on their heat scores.\\
\noindent \textit{Evaluation}: In the test set, we consider all the bugs posted in between July 2018 -- December 2018. Comments have been considered if it is posted in between July 2018 -- December 2019. This results in a total of 5302 bugs. Also, we have considered only the bugs which has at least one comment. Once again this ensures that each bug has one year of commenting time. Here, we use ground truth bug heat scores crawled in November 2020. The bugs in the test data are ranked based on these scores. We used the same test data across all the methods. \\
\noindent \textit{Graph setup}: While building the graph, we have considered only those packages which were reported to be affected in time periods reported above. The graph setup is transductive, i.e., we consider all the bugs in the training and the test data to construct the graph. Thus the graph consists of around 11137 nodes and $\sim 1205682$ edges. However, the model is made to observe the ground truth ranks of the nodes present in the training data only. The ground truth ranks for all the other nodes are hidden from the model.\\
\noindent \textit{Evaluation metrics}: In our experiments, we use three standard metrics to evaluate all our models. These are MAE (mean absolute error), MSE (mean squared error) and MAPE (mean average percentage error). Mean absolute error (MAE) is calculated as the average of absolute difference between the true scores and the predicted scores. Mean squared error (MSE) is calculated as the average of the squared differences between true scores and predicted scores. Mean average percentage error (MAPE) is calculated as the average of the absolute difference between ground truth and predicted values expressed as a percentage of the ground-truth value. \\
\noindent \textbf{Hyperparameters -- Text based approaches}: In text based approaches, we have run the models on 70\% training data. Further, we choose the text based model which performs best among all the text based models and compare it with the graph based models.\\
\noindent \textbf{Doc2Vec}:\label{doc2vec:setup} In our experiment, the input feature is a concatenation of the 100 dimensional vector representations of the description and the comments. The hyperparameters of the \mlp~for the regression task are set as follows. The learning rate, $\epsilon$ are set to $5e-4$, $1e-5$ respectively. The batch size and weight decay are set to 64 and 0.01 respectively. The number of neurons in hidden layer is 128 unit. We run the model for 20 epochs and saved the model where the current validation loss is better than the current best validation loss.\\
\noindent \textbf{SBERT}: Here, we have used the pretrained  \sbert~model called `paraphrase-mpnet-base-v2'. We have generated the embeddings for the description and comments of each bug separately. The dimension for each embedding is 768. The two embeddings are then concatenated to construct the feature.  For 70\% setup, the \mlp~learning rate, $\epsilon$, batch size and weight decay are set to $5e-3$, $2e-5$, 64 and 0.3 respectively. We have used one hidden layer with 1024 unit neurons and one output layer. For both the layers, we used \textsc{RelU} activation function. We observed that \sbert~performs better than Doc2Vec (see section~\ref{doc2vec:setup}). We therefore use \sbert~as the competing text-based baseline. Therefore we had to carry out experiments on \sbert~for other training data setup (50\% - 5\%). We use grid search to obtain the best parameter for the different training setups. For a 50\% training setup, the learning rate is $1e-3$. The weight decay and $\epsilon$ remains the same as that of the 70\% training setup. The hidden layer and batch size remains the same for all the setups, i.e., 1024, $2e-5$, and 64 respectively. The $\epsilon$ value of 50\% and 5\% setups is the same as 70\% setup. For 25\% and 10\% setups, the $\epsilon$ value is $1e-5$. In the 25\% and 5\% training data setup, the learning rate is $4e-3$. In the 50\% and 10\% training data setup, the learning rate is set to $1e-3$. For 25\%, 10\% and 5\%, the value of weight decay is 0.01. The weight decay value for the 50\% setup is the same as the 70\% setup.\\
\begin{table*}[!ht]
\centering
\scalebox{1.1}{
\begin{tabular}{|c|c|c|c|} \hline 
\multirow{2}{*}{\bf Methods} & \multicolumn{3}{c|}{\bf Text based}\\
\cline{2-4}
 & {\bf MAE} & {\bf MSE} & {\bf MAPE} \\ \hline \hline
Doc2Vec & 7.585 & 58.503 & $3.42e-16$ \\ \hline
\sbert & \textbf{1.081} &  \textbf{1.56} & \textbf{0.1649}\\ \hline
\end{tabular}
}
\caption{Results from the text based models for 70\% training setup. Best results are marked in boldface.}\label{tab:textbased}
\end{table*}
\noindent \textbf{Results from text based approaches}: The key results obtained from the text-based models are noted in Table~\ref{tab:textbased}. The results show that the \sbert~based model outperforms the other models. Hence we have used this model to compare the different graph-based approaches in the next section. The results of 50\%-5\% data setup for the \sbert~model are presented in Table~\ref{tab:main_results}.\\
\noindent \textbf{Hyperparameters -- Graph based approaches}: In our graph based approaches, we perform the experiments for different proportions of training data: 70\%, 50\%, 25\%, 10\%, 5\%. In each case, the rest of the data is used for validation.\\
\noindent \textbf{GAT}: In this experiment, we use Adam optimizer,  MAE, as the loss function and patience of early stopping at 15 in all the setups. We use grid search again to obtain the best parameters. For 70\% and 5\% setup, the attention head size is 32. For 25\% and 10\%, the size of the attention head is 64. In the case of 50\% setup, the attention head size remains 16. For 70\%, 50\% and 5\% training data, the \gat~layer sizes are 32, 16 and 32 respectively, and for other setups, it remains at 64. The best learning rate for 70\%, 25\% and 10\% training data setup are 0.01. The best learning rate for 50\% and 5\% training data is 0.015, respectively. The parameter in\_dropout is 0.5 for 50\% and 25\% training setup, respectively. For 70\%, 10\% and 5\% training data, the in\_dropouts are 0.9, 0.85 and 0.6 respectively. The attention dropout parameter is 0.1 for 10\% training setup. For other setups, it is 0.3. For all the setups, the activation function used in \gat~layer is \textsc{elU}.\\
\noindent \textbf{GCN}: For all the setup, we have used Adam optimizer, MAE as loss function, patience is 15 for early stopping and \textsc{elU} activation for \gcn~layer. We use grid search to obtain the best parameters. For 70\%, 25\% and 5\% setup, we have set the learning rate at 0.01. For 50\% and 10\% setup, the learning rate is set to 0.015 and 0.008, respectively. For all the setup except 5\% setup, \gcn~layer size is kept at 64. For the 5\% setup, the \gcn~layer size is 32. For 70\% and 5\% training setup, the dropout value is set at 0.85. For 25\% and 10\% training setup, the dropout is set to 0.8. For 50\% setup, the dropout is 0.9.\\
\noindent \textbf{GRAPHSAGE}: For all the setup, we use Adam optimizer, MAE as loss function and the early stopping patience at 15. Activation function used in \sage~layer is \textsc{elU}. In the last layer, we have used \textsc{RelU} activation function, and bias is set to true. The batch size is set to 64. For each training proportion, we use grid search to obtain the best parameter setting. For 70\% setup, the size of the \sage~layer is 64, and for other setups, the size of the \sage~layer is 128. For 70\% and 10\%, the learning rate is $1e-2$. For 50\%, 25\% and 5\%, the learning rate is $15e-3$. For 70\% and 5\% setup, the number of nodes sampled in each \sage~layer is 5. For 50\%, 25\% and 10\% setup, the number of nodes sampled in \sage~layer is 6. For 50\% setup, the dropout in \sage~layer are 0.6 and for other setups the dropout in \sage~layer is 0.9.\\
\begin{table*}[!ht]
\scalebox{0.8}{
\begin{tabular}{c|c|c|c||c|c|c|c|c|c|c|c|c} \hline
\multirow{3}{*}{\bf Training data}&\multicolumn{3}{c||}{Text based} & \multicolumn{9}{c}{Graph based}\\ 
\cline{2-13}

 & \multicolumn{3}{c||}{\sbert} & \multicolumn{3}{c|}{\gat} & \multicolumn{3}{c|}{\gcn} & \multicolumn{3}{c}{\sage}  \\ 
 \cline{2-13}
 & {\bf MAE} & {\bf MSE} & {\bf MAPE} & {\bf MAE} & {\bf MSE} & {\bf MAPE} & {\bf MAE} & {\bf MSE} & {\bf MAPE} & {\bf MAE} & {\bf MSE} & {\bf MAPE} \\ \hline

 \textbf{70\%}  & {\bf1.081}	 &  {\bf 1.56} & 	{\bf 0.1649} & 1.151 & 1.722 & 0.1736 &1.183 & 1.83 & 0.1804 &  \underline{1.148}  & 	\underline{1.796} & 	\underline{0.1752} \\
\cline{1-13}
 \textbf{50\%} & 0.908 & 1.19 & 0.1334 & \underline{0.812} & \underline{0.998} & \underline{	0.1131} &0.831 & 1.038 & 0.1165 & {\bf 0.795} & {\bf	0.94 } & {\bf 	0.1129}\\
\cline{1-13}
\textbf{25\%} & 1.069 & 1.688 & 0.1622 & {\bf 0.757}  & {\bf 	1.11} & {\bf 	0.1007} & \underline{0.833} & \underline{1.256} & \underline{0.1148} & 1.062 & 	1.436 & 0.1569 \\ 
\cline{1-13}
\textbf{10\%} &1.282 & 	2.565 & 0.2070 & {\bf 0.812	} & {\bf 1.124	} & {\bf 0.1096} & \underline{1.034} & \underline{	1.885} & \underline{0.1461} & 1.756 & 	3.658 & 	0.3001\\ 
\cline{1-13}
\textbf{5\%} & 1.178 & 	2.268 & 0.1778 & {\bf 0.849} & {\bf 1.415} & {\bf 0.1101} & \underline{1.089} & \underline{	2.09} & \underline{0.1532} & 2.242  & 5.721 & 0.4213\\ \hline
\end{tabular}
}
\caption{Results of the bug severity prediction using graph based methods. For each node the initial feature is a concatenation of the \sbert~embeddings of the {\bf description} and the {\bf comments}. The best results are indicated in boldface and second best are underlined.} 
\label{tab:main_results}
\end{table*}
\noindent \textbf{Results from graph based approaches}: Table~\ref{tab:main_results} summarises the main results of this section. We observe that for larger training data setup (70\% and 50\%), the text based model performs better than \gcn~and \gat. However, in this setting \sage~performs the second best in terms of all the evaluation metrics. For the low training data setup (5\% -- 25\%), the graph based models \gcn~and \gat~outperform the text based model. In fact, in this setup, \gat~performs the best in terms of all the evaluation metrics. This shows that enabling self-attention on the neighbourhood of a bug in the bug-bug network could be very beneficial for predicting bug severity when the number of training data points are severely low.
\section{Ablation study}
In our experiments, we have used a concatenation of representations of both the description and the comments. This section investigates the importance of each of these separately for the text and graph-based methods. The results obtained by using only the description are reported in Table~\ref{tab:ablationdesc}. The results for both the text-based and graph-based models obtained using only the comments are reported in Table~\ref{tab:ablationcomm}. \\
\noindent \textbf{Only description}:\label{ablation:onlydesc} We execute the text-based model (\sbert) and graph-based models (\gat~, \gcn~, \sage~) on only the description feature. In both types of models, we have used the text of the description as a feature. Further, we compute the results for all the training data setups (70\%-5\%). We wanted to observe whether one can obtain the additional support from graph information even if (s)he uses only one feature (i.e., description in this case).\\
\noindent\textit{Observations}: Out of all the setups, \sbert~has performed slightly better than one of the graph-based models in the 70\% setup. However, as the training data reduces, the performance of \sbert~ drops. For the 50\% setup, \sage~ model outperforms the other models, and \gat~ is the second best. However, the $MAE$ score difference between the best model (\sage~) and the \sbert~ is negligible ($\sim0.09$). For 25\% setup, \sage~ is again the top performer and \gat~is the second best. The difference in $MAE$ scores of \sbert~ and \sage~ models now are pretty large ($\sim0.27$). For the 10\% setup, \gat~ model performs better than other models. The $MAE$ score difference between \sbert~ and \gat~ is as large as 0.38. For 5\% setup, once again \gat~ performs better than other models. Here the second-best model is ~\gcn. ~It is visible that if the training data is reduced, then the text-based model does not perform well as was also observed in our original results (see Table~\ref{tab:main_results}). Nevertheless, the additional graph structure helps the model to predict better in a low training data setup. 
\begin{table*}[!h]
\scalebox{0.8}{
\begin{tabular}{c|c|c|c||c|c|c|c|c|c|c|c|c} \hline
\multirow{3}{*}{\bf Training data}&\multicolumn{3}{c||}{Text based} & \multicolumn{9}{c}{Graph based}\\ 
\cline{2-13}

 & \multicolumn{3}{c||}{\sbert} & \multicolumn{3}{c|}{\gat} & \multicolumn{3}{c|}{\gcn} & \multicolumn{3}{c}{\sage}  \\ 
 \cline{2-13}
 & {\bf MAE} & {\bf MSE} & {\bf MAPE} & {\bf MAE} & {\bf MSE} & {\bf MAPE} & {\bf MAE} & {\bf MSE} & {\bf MAPE} & {\bf MAE} & {\bf MSE} & {\bf MAPE} \\ \hline

 \textbf{70\%}  & \textbf{1.035}	 & \textbf{1.474} & \textbf{0.1530} & 1.152 & 1.735 & 0.1738 &  1.194 & 1.931 & 0.1841 &  \underline{1.093} & 	\underline{1.629} & \underline{0.1635} \\
\cline{1-13}
 \textbf{50\%} &  0.905 & 1.205 & 0.1294 & \underline{0.821} & \underline{1.03} & \underline{0.114} &  0.85 & 1.081 & 0.1193 &  \textbf{0.812}	& \textbf{1.019} & \textbf{0.1125}\\
\cline{1-13}
\textbf{25\%} &  0.987 & 1.555 & 0.144  & \underline{0.784}  & 	\underline{1.119} & \underline{0.1049} &  0.934 & 1.465 & 0.1341 &  \textbf{0.717} & 	\textbf{0.954} & \textbf{0.0949} \\ 
\cline{1-13}
\textbf{10\%} &  1.275 & 2.458 & 0.2082 & \textbf{0.892}	& \textbf{1.607}	 & \textbf{0.1164} &   \underline{1.067} & 	\underline{2.026} & \underline{0.154} &  1.677 & 3.286 & 0.2782\\ 
\cline{1-13}
\textbf{5\%} &  1.183 & 2.401 & 0.1731 & \textbf{0.87} & \textbf{1.54} & \textbf{0.1153} & \underline{1.057} & \underline{2.004} & \underline{0.1531} &  2.213 & 	5.566 & 	0.412 \\ \hline
\end{tabular}
}
\caption{{\bf Ablation study}: MAE, MSE and RMSE values are reported to compare best text based methods with different graph based methods. Only {\bf description} of the bugs are used to generate the embedding.}
\label{tab:ablationdesc}
\end{table*}
\begin{table*}[!ht]
\scalebox{0.8}{
\begin{tabular}{c|c|c|c||c|c|c|c|c|c|c|c|c} \hline
\multirow{3}{*}{\bf Training data}&\multicolumn{3}{c||}{Text based} & \multicolumn{9}{c}{Graph based}\\ 
\cline{2-13}

 & \multicolumn{3}{c||}{\sbert} & \multicolumn{3}{c|}{\gat} & \multicolumn{3}{c|}{\gcn} & \multicolumn{3}{c}{\sage}  \\ 
 \cline{2-13}
 & {\bf MAE} & {\bf MSE} & {\bf MAPE} & {\bf MAE} & {\bf MSE} & {\bf MAPE} & {\bf MAE} & {\bf MSE} & {\bf MAPE} & {\bf MAE} & {\bf MSE} & {\bf MAPE} \\ \hline
 \textbf{70\%}  &  \textbf{0.958} & \textbf{1.298} & \textbf{0.1414} & \underline{1.103} & \underline{1.621} & \underline{0.1651} &  1.195 & 1.89	 & 0.1838 &  1.107 & 	1.66 & 	0.1660 \\
\cline{1-13}
 \textbf{50\%} &  0.908 & 1.196 & 0.1337 & \underline{0.802} & \underline{0.984} & \underline{0.1109} &  0.806 & 	0.996 & 0.112 &  \textbf{0.782} &	\textbf{0.925}&	\textbf{0.1113}  \\
\cline{1-13}
\textbf{25\%} & 1.018 & 1.677 & 0.1537 &  \textbf{0.743}	& \textbf{1.044} & \textbf{0.0984} &   \underline{0.871}	& \underline{1.334} & \underline{0.1226}  &  1.081 &	1.489 &	0.1603 \\ 
\cline{1-13}
\textbf{10\%} &   1.431 & 3.35 & 0.2348 & \textbf{0.924}  &	\textbf{1.322} & \textbf{0.1305} &  \underline{1.122} & \underline{2.128} & \underline{0.1682} &  1.724 &	3.518 &	0.2916 \\ 
\cline{1-13}
\textbf{5\%} &  1.485 & 3.601 & 0.2618 & \textbf{1.004} & \textbf{1.809} & \textbf{0.1420} & \underline{1.171} & \underline{2.244} & \underline{0.1794}  &  2.205 &	5.531	& 0.4099 \\ \hline
\end{tabular}
}
\caption{{\bf Ablation study}: MAE, MSE and RMSE values are reported to compare best text based methods with different graph based methods. Only {\bf comments} of the bugs are used to generate the embedding.} 
\label{tab:ablationcomm}
\end{table*}
\noindent \textbf{Only comments}:\label{subsec:onlycomments} Here we carry out the experiments using only one feature, i.e., comments. Once again the idea is to verify whether the graph structure is useful even when one of the feature types are available. We have taken the texts of the comments in this experiment. We ran the experiments for all the text-based and graph-based models. We perform these experiments for all the training setups (70\%-5\% training data). \\
\noindent\textit{Observations}: For 70\% training setup, like in the previous experiment (subsection~\ref{ablation:onlydesc}), \sbert~ is performing well than other graph-based models. But the $MAE$ score difference between the \sbert ~and best performing graph model (\gat~) is quite less ($\sim0.145$). For the 50\% training setup, all the three graph models are performing better than the \sbert~ model. Out of three graph models, \sage~ performed best ($MAE$ 0.782 ), and \sbert is performs the worst ($MAE$ 0.908). For the 25\% setup, the \gat~ model outperforms other models, and the $MAE$ difference between the \sbert and \gat model is quite high ($\sim0.275$). Here, the \sage~ performs the worst, and \sbert~ performs better than the \sage model. For 10\% setup, again \gat~model tops the list and \gcn~comes as second best. For 5\% setup, \gat~ outperforms other models, and \gcn~ is the second in the list. Overall once again we observe that the graph structure is always helpful whatever be the text feature especially in the low data setting.
\section{Error analysis}
\label{sec:erroranalysis}
In this section, we will test our models for various cases and identify which models fail when and why. First, we shall test the importance of the graph structure. Second, we shall study some cases where \sbert~fails, but \gat~wins and vice versa. From the description+comments results, we observe that for low data, the \sage~performs poorer than \gat~ (best) model. Hence we shall consider some use cases to analyze this fact first.\\
\noindent \textbf{Testing the importance of graph structure}: We perform error analysis for low training data setup (5\%-25\%) to understand the importance of the graph structure. Among all the models, \gat~performs better for low training data setup. In order to carry out the analysis, given a model, we first calculate the absolute difference (i.e., $\Delta$) between the predicted rank and the true rank of the bugs. Further, we compute the number of bugs where absolute difference ($\Delta$\gat) in \gat~is lesser (i.e., better) than ($\Delta$\sbert) in \sbert. The results of this analysis are summarised in  Table~\ref{tab:errorgraph}. As we can observe, for 5\% data, in 62.69\% of test cases, $\Delta$\gat~$<$ $\Delta$\sbert. Out of these, 60.10\% of the test bugs have a neighbor in the bug-bug network that was part of the training node of \gat. Similar results hold for the other cases. This shows that the graph structure indeed helps in improving the predicted ranks in low data settings. 
\begin{table*}[!htb]
\centering
\scalebox{1}{
\begin{tabular}{c|c|c|c} \hline 
 Training data &  $\Delta$\gat~$<$ $\Delta$\sbert & $\Delta$\gat~$<$ $\Delta$\sbert &  \#nodes in training \\ 
 &&(has neighbor in training)& \\
  \hline \hline
{\bf 25\%} & 3485 (65.72\%) & 2609 (74.86\%) & 1458\\ \hline 
{\bf 10\%} & 3712 (70.01\%) & 2359 (63.55\%)  & 583 \\ \hline
{\bf 5\%} & 3324 (62.69\%) &  1998 (60.10\%) &291\\ \hline
\end{tabular}
}
\caption{\label{tab:errorgraph}Outcomes of the error analysis. Results are only shown for the low data setup to investigate the importance of the graph neighborhood.}
\end{table*}

\noindent \textbf{Usecase: \gat~wins \sbert ~fails}: In Table ~\ref{tab:gatbetterthansbert}, we present a few example test bugs where the \gat ~model predicts a better rank value than the \sbert~ model. For this analysis, we use the prediction value from the models trained with 5\% training data. We have chosen 5\% training data because, especially for low data, the graph-based method outperforms the text-based model. For each of the test bugs, we calculate the number of neighbours present in training set. We observe that in case of test bugs with a large number of neighbours in the training set for \gat~predicts nearer ranks to the ground truth rank compared to \sbert~. In these examples, most of the test bugs have 27-29 neighbouring nodes in training data (note: the total number of nodes in training data is 291).

\begin{table*}[!ht]
\centering
\scalebox{1}{
\begin{tabular}{c|c|c|c|c|c|c} \hline 
Bug Id & True rank & Prediction \sbert & Prediction \gat & $\Delta$\sbert & $\Delta$\gat & \#neighbors \\ 
 &&&&&&( in training ) \\
  \hline \hline
1799406 & 7.669 & 8.450 & 8.050	& 0.780 & 0.380 & 29 \\ \hline 
1792783	& 7.920 & 5.804 & 8.048 & 2.116 & 0.128 & 28\\ \hline 
1798690	& 7.920 & 6.086 & 8.251 & 1.833	& 0.330 & 28 \\ \hline
1788045 & 8.338 & 7.354 & 8.252 & 0.983	& 0.085 & 27\\ \hline
\end{tabular}
}
\caption{\label{tab:gatbetterthansbert} Few test examples where the \gat~ model predicts a nearer value to the true rank compared to \sbert. ~In all these cases the instances have a lot of neighbors present in training data.}
\end{table*}
\noindent \textbf{Usecase: \sbert~wins \gat~fails}: Converse to the data points in the previous section, here we find that there are a set of test points for which \gat ~fails even in the low (i.e., 5\%) training data setup, i.e., the ranks predicted by \gat~ are further from the ground truth compared to \sbert.~ In all these cases we observe that the number of nodes in the training set for each of these test points is 0 (see Table~\ref{tab:sbertbetterthangat}). The absence of neighbours of these points in the training set does not allow the \gat~model to take the advantage of the graph structure and hence the worse rank. 

\begin{table*}[!ht]
\centering
\scalebox{1}{
\begin{tabular}{c|c|c|c|c|c|c} \hline 
Bug Id & True rank & Prediction \sbert & Prediction \gat & $\Delta$\sbert & $\Delta$\gat & \#neighbors\\ 
 &&&&&&( in training ) \\
  \hline \hline
1797179 & 7.397 & 8.332 & 9.005 & 0.934 & 1.607 & 0 \\ \hline 
1788706 & 6.772 & 7.881 & 8.511 & 1.108	& 1.738 & 0\\ \hline 
1810154 & 6.050 & 7.127 & 7.759 & 1.076	& 1.708 & 0 \\ \hline
1791333 & 8.338 & 8.556 & 7.509 & 0.218 & 0.828 & 0\\ \hline
\end{tabular}
}
\caption{\label{tab:sbertbetterthangat} Few test examples where the \sbert~ model predicts a nearer value to the true rank compared to \gat. ~In all these cases the instances have 0 neighbors present in training data.}
\end{table*}
\begin{table*}[!ht]
\centering
\scalebox{0.86}{
\begin{tabular}{c|c|c|c|c|c|c} \hline 
Bug Id & True rank & Prediction \sage & Prediction \gat & $\Delta$\sage & $\Delta$\gat & \#neighbors \\ 
 &&&&&&( in training ) \\
  \hline \hline
1799406 & 7.669 & 5.609 & 8.050	& 2.060 & 0.380 & 29 \\ \hline 
1793137	& 8.338 & 5.564 & 8.055 & 2.773  & 0.282 & 28\\ \hline 
1798690	& 7.920 & 5.614 & 8.251 & 2.305	& 0.330 & 28 \\ \hline
1788045 & 8.338 & 5.619 & 8.252 & 2.718	& 0.085 & 27\\ \hline
\end{tabular}
}
\caption{\label{tab:gatbetterthansage}Test examples where the \gat~ model predicts ranks closer to the ground truth compared to the \sage~ model.}
\end{table*}
\noindent \textbf{Usecase: \gat~wins \sage~fails}: In Table~\ref{tab:gatbetterthansage}, we list a few test cases where the \gat~wins but \sage~fails for low data setup (5\%). We list those cases where there is a sufficient number of neighbours in training data, but still, the performance of \sage~ is poor. The $\Delta$\gat~ is much lesser than the $\Delta$\sage~ for all the cases. We pick up every instance and try to understand the 1.5 hop neighbourhood structure. We focus on those neighbours specifically who are present in the training set. So, for each test bug (say the anchor node), we build a 1.5 neighbourhood (taking only those neighbours which are present in the training data) graph. 1.5 neighbourhood graph contains the anchor node and its neighbours and connection among themselves (i.e., anchor to neighbours as well as among neighbours). We observe that the degree centrality of the neighbours vary. Thus one can hypothesize that all the neighbours (present in training set) are not equally important for the prediction of the rank of the anchor node. In the \gat ~architecture, the model provides different attention coefficients to different neighbours of each anchor node to compute the representation of the node. Also, the feature aggregation has been done based on the importance (attention coefficient) of immediate neighbours of the anchor node. However, in case of the \sage~ model, a certain number of nodes has been uniformly sampled from the set of neighbours. Further, the feature aggregation for each anchor node is done based on the sampled neighbourhood. Using the differences in the attention coefficients the \gat~ model possibly leverages more information from the high degree neighbors in order to predict the rank of the anchor node thus outperforming the \sage~model which gives uniform importance to all the neighbors of the anchor node.
\section{Conclusion}
\label{conc}
In this paper, we presented a new dataset comprising bugs, its metadata and ground truth severity scores (i.e., bug heat) from two time points. Further, we collected the list of affected packages by a bug along with the timestamp. We build regression models for bug severity prediction which is one of the well known problems in the software community. We performed the experiments using two type of models -- (i) text based models, and (ii) graph based models. We observed that the \sbert~model performed better (/similar) for high training data (70\%, 50\%) setup than graph based models. However, for low training data setup, the \gat~model outperformed \sbert~by a large margin. Error analysis shows that the performance of \gat~is due to the nodes in the training set of the model that are in the neighbourhood of the bug-bug network of the test bugs. In future, we would like to carry out the studies such as how the packages are being affected temporally using our new dataset.

%
%
%

\begin{thebibliography}{8}
\bibitem{Umer:2018}
Q. {Umer} and H. {Liu} and Y. {Sultan}.: Emotion Based Automated Priority Prediction for Bug Reports. IEEE Access, vol. 6, pp. 35743--35752 (2018).

\bibitem{Tan:2020}
Youshuai Tan and Sijie Xu and Zhaowei Wang and Tao Zhang and Zhou Xu and Xiapu Luo.:Bug severity prediction using question-and-answer pairs from Stack Overflow. Journal of Systems and Software, vol. 165, pp. 110567 (2020).

\bibitem{Arokiam:2020}
Arokiam, Jude and Bradbury, Jeremy S.:Automatically Predicting Bug Severity Early in the Development Process. Proceedings of the ACM/IEEE 42nd International Conference on Software Engineering: New Ideas and Emerging Results, pp. 17–20 (2020). 

\bibitem{Wu:2021}
Xiaoxue Wu and Wei Zheng and Xiang Chen and Yu Zhao and Tingting Yu and Dejun Mu.: Improving high-impact bug report prediction with combination of interactive machine learning and active learning. Information and Software Technology, vol. 133, pp. 106530 (2021). 

\bibitem{Ohira:2015}
Ohira, Masao and Kashiwa, Yutaro and Yamatani, Yosuke and Yoshiyuki, Hayato and Maeda, Yoshiya and Limsettho, Nachai and Fujino, Keisuke and Hata, Hideaki and Ihara, Akinori and Matsumoto, Kenichi.: A Dataset of High Impact Bugs: Manually-Classified Issue Reports. 2015 IEEE/ACM 12th Working Conference on Mining Software Repositories, pp. 518-521 (2015). 

\bibitem{Wu:2020}
Xiaoxue Wu and Wei Zheng and Xiang Chen and Fang Wang and Dejun Mu.: CVE-assisted large-scale security bug report dataset construction method. Journal of Systems and Software, vol. 160, pp. 110456 (2020).

\bibitem{Peters:2019}
Peters, Fayola and Tun, Thein Than and Yu, Yijun and Nuseibeh, Bashar.: Text Filtering and Ranking for Security Bug Report Prediction. IEEE Transactions on Software Engineering, vol. 45, pp. 615-631 (2019).

\bibitem{Ramay:2019}
Ramay, Waheed Yousuf and Umer, Qasim and Yin, Xu Cheng and Zhu, Chao and Illahi, Inam.: Deep Neural Network-Based Severity Prediction of Bug Reports. IEEE Access, vol. 7, pp. 46846-46857 (2019).  

\bibitem{Lamkanfi:2013}
Lamkanfi, Ahmed and P\'{e}rez, Javier and Demeyer, Serge.: The Eclipse and Mozilla Defect Tracking Dataset: A Genuine Dataset for Mining Bug Information. Proceedings of the 10th Working Conference on Mining Software Repositories, pp. 203-206 (2013).

\bibitem{mikolov:2013}
Mikolov, Tomas and Chen, Kai and Corrado, Greg and Dean, Jeffrey.: Efficient Estimation of Word Representations in Vector Space. 1st International Conference on Learning Representations (ICLR) (2013).

\bibitem{Goseva-Popstojanova:2018}
Goseva-Popstojanova, Katerina and Tyo, Jacob.: Identification of Security Related Bug Reports via Text Mining Using Supervised and Unsupervised Classification. 2018 IEEE International Conference on Software Quality, Reliability and Security (QRS), pp. 344-355 (2018).

\bibitem{Hazra:2021}
Hazra, Rima and Aggarwal, Hardik and Goyal, Pawan and Mukherjee, Animesh and Chakrabarti, Soumen.: Joint Autoregressive and Graph Models for Software and Developer Social Networks. Advances in  Information Retrieval (ECIR), pp. 224--237 (2021).

\bibitem{quoc:2014}
Le, Quoc and Mikolov, Tomas.: Distributed Representations of Sentences and Documents. Proceedings of the 31st International Conference on Machine Learning, vol. 32, pp. 1188--1196 (2014).

\bibitem{reimers:2019}
Reimers, Nils and Gurevych, Iryna.: Sentence-BERT: Sentence Embeddings using Siamese BERT-Networks. Proceedings of the 2019 Conference on Empirical Methods in Natural Language Processing (EMNLP) (2019).

\bibitem{Devlin:2019}
Jacob Devlin and Ming-Wei Chang and Kenton Lee and Kristina Toutanova.: BERT: Pre-training of Deep Bidirectional Transformers for Language Understanding. Proceedings of the 2019 Conference of the North {A}merican Chapter of the Association for Computational Linguistics: Human Language Technologies, vol. 1, pp. 4171--4186 (2019).

\bibitem{velickovic:2018}
Veli{\v{c}}kovi{\'{c}}, Petar and Cucurull, Guillem and Casanova, Arantxa and Romero, Adriana and Li{\`{o}}, Pietro and Bengio, Yoshua.: Graph Attention Networks. International Conference on Learning Representations (2018).

\bibitem{kipf:2017}
Kipf, Thomas N. and Welling, Max.: Semi-Supervised Classification with Graph Convolutional Networks. International Conference on Learning Representations (ICLR) (2017).

\bibitem{hamilton:2017}
Hamilton, William L. and Ying, Rex and Leskovec, Jure. Inductive Representation Learning on Large Graphs. NIPS (2017).





\end{thebibliography}
%

\end{document}